\documentclass[aps,prl,twocolumn,showpacs,superscriptaddress,amsmath,amssymb]{revtex4}

\usepackage{graphicx}
\usepackage{dcolumn}
\usepackage{bm}
\begin{document}


\title{On the extraction of weak transition strengths via the ($^{3}$He,$t$) reaction at 420 MeV.}

\author{R.G.T. Zegers}
\affiliation{National Superconducting Cyclotron Laboratory, Michigan
State University, East Lansing, MI 48824-1321, USA}
\affiliation{Department of Physics and Astronomy, Michigan State
University, East Lansing, MI 48824, USA} \affiliation{Joint
Institute for Nuclear Astrophysics, Michigan State University, East
Lansing, MI 48824, USA}
\author{T. Adachi}
\affiliation{Research Center for Nuclear Physics, Osaka University,
Ibaraki, Osaka 567-0047, Japan}
\author{H. Akimune}
\affiliation{Department of Physics, Konan University, Kobe, Hyogo, 658-8501, Japan}
\author{Sam M. Austin}
\affiliation{National Superconducting Cyclotron Laboratory, Michigan
State University, East Lansing, MI 48824-1321, USA}
\affiliation{Joint Institute for Nuclear Astrophysics, Michigan
State University, East Lansing, MI 48824, USA}
\author{A.M. van den Berg}
\affiliation{Kernfysisch Versneller Instituut, University of
Groningen, Zernikelaan 25, 9747 AA Groningen, The Netherlands}
\author{B.A. Brown}
\affiliation{National Superconducting Cyclotron Laboratory, Michigan State
University, East Lansing, MI 48824-1321, USA} \affiliation{Department of Physics and
Astronomy, Michigan State University, East Lansing, MI 48824, USA}
\affiliation{Joint Institute for Nuclear Astrophysics, Michigan State University,
East Lansing, MI 48824, USA}
\author{Y. Fujita}
\affiliation{Department of Physics, Osaka University, Toyonaka, Osaka 560-0043, Japan}
\author{M. Fujiwara}
\affiliation{Kansai Photon Science Institute, Japan Atomic Energy Agency, Kizu, Kyoto 619-0215, Japan}
\affiliation{Research Center for Nuclear Physics, Osaka University, Ibaraki, Osaka 567-0047, Japan}
\author{S. Gal\`{e}s}
\affiliation{Institut de Physique Nucl\'{e}aire, IN2P3-CNRS, Orsay, France}
\author{C.J. Guess}
\affiliation{National Superconducting Cyclotron Laboratory, Michigan State
University, East Lansing, MI 48824-1321, USA} \affiliation{Department of Physics and
Astronomy, Michigan State University, East Lansing, MI 48824, USA}
\affiliation{Joint Institute for Nuclear Astrophysics, Michigan State University,
East Lansing, MI 48824, USA}
\author{M.N. Harakeh}
\affiliation{Kernfysisch Versneller Instituut, University of Groningen, Zernikelaan 25, 9747 AA Groningen, The Netherlands}%
\author{H. Hashimoto}
\affiliation{Research Center for Nuclear Physics, Osaka University, Ibaraki, Osaka 567-0047, Japan}
\author{K. Hatanaka}
\affiliation{Research Center for Nuclear Physics, Osaka University,
Ibaraki, Osaka 567-0047, Japan}
\author{R. Hayami}
\affiliation{Department of Physics, University of Tokushima, Tokushima 770-8502, Japan}
\author{G.W. Hitt}
\affiliation{National Superconducting Cyclotron Laboratory, Michigan State
University, East Lansing, MI 48824-1321, USA} \affiliation{Department of Physics and
Astronomy, Michigan State University, East Lansing, MI 48824, USA}
\affiliation{Joint Institute for Nuclear Astrophysics, Michigan State University,
East Lansing, MI 48824, USA}
\author{M.E. Howard}
\affiliation{Joint Institute for Nuclear Astrophysics, Michigan State University, East Lansing, MI 48824, USA}
\affiliation{Department of Physics, The Ohio State University, Columbus, OH 43210, USA}
\author{M. Itoh}
\affiliation{Cyclotron and Radioisotope Center, Tohoku University,
Sendai, Miyagi 980-8578, Japan}
\author{T. Kawabata}
\affiliation{Center for Nuclear Study, University of Tokyo, RIKEN Campus, Wako, Saitama 351-0198, Japan}
\author{K. Kawase}
\affiliation{Kansai Photon Science Institute, Japan Atomic Energy Agency, Kizu,
Kyoto 619-0215, Japan}
\author{M. Kinoshita}
\affiliation{Department of Physics, Konan University, Kobe, Hyogo, 658-8501, Japan}
\author{M. Matsubara}
\affiliation{Research Center for Nuclear Physics, Osaka University,
Ibaraki, Osaka 567-0047, Japan}
\author{K. Nakanishi}
\affiliation{Research Center for Nuclear
Physics, Osaka University, Ibaraki, Osaka 567-0047, Japan}
\author{S. Nakayama}
\affiliation{Department of Physics, University of Tokushima, Tokushima 770-8502, Japan}
\author{S. Okumura}
\affiliation{Research Center for Nuclear Physics, Osaka University,
Ibaraki, Osaka 567-0047, Japan}
\author{T. Ohta}
\affiliation{Research Center for Nuclear Physics, Osaka University,
Ibaraki, Osaka 567-0047, Japan}
\author{Y. Sakemi}
\affiliation{Research Center for Nuclear Physics, Osaka University,
Ibaraki, Osaka 567-0047, Japan}
\author{Y. Shimbara}
 \affiliation{National Superconducting Cyclotron
Laboratory, Michigan State University, East Lansing, MI 48824-1321,
USA}
\author{Y. Shimizu}
\affiliation{Research Center for Nuclear Physics, Osaka University,
Ibaraki, Osaka 567-0047, Japan}
\author{C. Scholl}
\affiliation{Institut f\"ur Kernphysik, Universit\"at zu K\"oln, D-50937 Cologne, Germany}
\author{C. Simenel}
\affiliation{National Superconducting Cyclotron Laboratory, Michigan State University, East
Lansing, MI 48824-1321, USA} \affiliation{CEA/DSM/DAPNIA/SPhN, F-91191 Gif-sur-Yvette, France}
\author{Y. Tameshige}
\affiliation{Research Center for Nuclear Physics, Osaka University,
Ibaraki, Osaka 567-0047, Japan}
\author{A. Tamii}
\affiliation{Research Center for Nuclear Physics, Osaka University,
Ibaraki, Osaka 567-0047, Japan}
\author{M. Uchida}
\affiliation{Tokyo Institute of Technology, 2-12-1 O-Okayama, Tokyo
152-8550, Japan}
\author{T. Yamagata}
\affiliation{Department of Physics, Konan University, Kobe, Hyogo, 658-8501, Japan}
\author{M. Yosoi}
\affiliation{Research Center for Nuclear Physics, Osaka University, Ibaraki, Osaka 567-0047, Japan}
\date{\today}%

\begin{abstract}
Differential cross sections for transitions of known weak strength were measured with the
($^{3}$He,$t$) reaction at 420 MeV on targets of $^{12}$C, $^{13}$C, $^{18}$O, $^{26}$Mg,
$^{58}$Ni, $^{60}$Ni, $^{90}$Zr, $^{118}$Sn, $^{120}$Sn and $^{208}$Pb. Using these data, it is
shown the proportionalities between strengths and cross sections for this probe follow simple
trends as a function of mass number. These trends can be used to confidently determine Gamow-Teller
strength distributions in nuclei for which the proportionality cannot be calibrated via
$\beta$-decay strengths. Although theoretical calculations in distorted-wave Born approximation
overestimate the data, they allow one to understand the main experimental features and to predict
deviations from the simple trends observed in some of the transitions.
\end{abstract}

\maketitle

\section{Introduction}
\label{sec:intro} The ($^{3}$He,t) charge-exchange (CE) reaction is widely used to study the
spin-isospin response of nuclei \cite{FUJ96,HAR01}. In particular, Gamow-Teller transitions (GT;
transfer of spin $\Delta$S=1, of orbital angular momentum $\Delta$L=0 and of total angular momentum
$\Delta$J=1) have been the subject of intensive investigations, since they are used to extract weak
transition strengths in excitation-energy regions inaccessible to $\beta$-decay. These strength
distributions are crucial for understanding such diverse topics as late stellar evolution
\cite{FUJ05,ADA06}, neutrino nucleosynthesis \cite{BYE07}, design of neutrino detectors
\cite{FUW00} and for constraining calculations of matrix elements for (neutrinoless) double
$\beta$-decay \cite{AKI97,EJI05}.

Compared to the $(p,n)$ reaction, use of the ($^{3}$He,$t$) reaction at intermediate energies has
the distinct advantage that much better energy resolutions (as low as 20 keV \cite{FUJ07}) can be
achieved. This permits a cleaner extraction of GT strengths (B(GT)) from competing transitions and
provides a higher level of detail. However, unlike for the $(p,n)$ reaction, there has been no
convincing systematic evaluation of the reliability of extracting B(GT) values using the
($^{3}$He,$t$) reaction. Such an evaluation is crucial for obtaining reliable inputs for the
above-mentioned applications and for use of the inverse reaction (t,$^{3}$He) to extract electron
capture strengths \cite{HIT06,COL06}.

In this letter, we fill this gap by measuring the differential cross sections for the
($^{3}$He,$t$) reaction at E($^{3}$He)=420 MeV on target nuclei over a wide mass range. The data
are used to study the ($^{3}$He,$t$) reaction mechanism and procedures for extracting B(GT). We
find, and understand, differences from the $(p,n)$ reaction and show that the accuracy of the
extracted B(GT) values are comparable for the two probes.

The extraction of weak transition strengths from CE data is based on the close
proportionality between the weak transition strength and the CE differential cross
section at zero momentum transfer (($\frac{\text{d}\sigma}{\text{d}\Omega}(q=0$))
derived in the Eikonal approximation \cite{TAD87}. For GT transitions:
\begin{equation}
\label{eq:eik}
\frac{d\sigma}{d\Omega}(q=0)=KN_{\sigma\tau}|J_{\sigma\tau}|^{2}B(GT)=\hat{\sigma}_{GT}B(GT).
\end{equation}
Here, $K$ is a kinematical factor, $N_{\sigma\tau}$ a distortion factor defined by
the ratio of distorted-wave to plane-wave cross sections, and $|J_{\sigma\tau}|$ is
the volume-integral of the central $\sigma\tau$ interaction. The factor
$KN_{\sigma\tau}|J_{\sigma\tau}|^{2}$ is referred to as the unit cross section,
$\hat{\sigma}_{GT}$. For Fermi transitions ($\Delta$S=0, $\Delta$L=0, $\Delta$J=0),
$|J_{\sigma\tau}|$ has to be replaced by $|J_{\tau}|$, $B(GT)$ by $B(F)$ and
$N_{\sigma\tau}$ by $N_{\tau}$ and the unit cross section is referred to as
$\hat{\sigma}_{F}$. The validity of Eq. \ref{eq:eik} was studied for the $(p,n)$
reaction \cite{TAD87} on a wide variety of target nuclei at beam energies above
$E_{p}$=120 MeV. This study made use of transitions for which the B(GT) values are
known from $\beta$-decay experiments. For Fermi transitions, the sum-rule strength
($B(F)=N-Z$) is nearly exhausted by the excitation of the Isobaric Analog State
(IAS) \cite{IKE63}.

The ($^{3}$He,$t$) and $(p,n)$ reactions differ significantly. The $^{3}$He ($t$) has internal
structure and is absorbed at the surface of the target nucleus, whereas the proton (neutron) is a
single nucleon that probes the nuclear interior. Differences between the two probes have become
apparent in experimental studies of the ratio $\frac{\hat{\sigma}_{GT}}{\hat{\sigma}_{F}}$. Whereas
it is nearly independent of mass number for the $(p,n)$ reaction \cite{TAD87}, a significant
increase has been observed for the ($^{3}$He,$t$) reaction \cite{ADA07} that, until now, was poorly
understood. Such issues have led to concerns about the validity of Eq. \ref{eq:eik} for the
($^{3}$He,$t$) reaction at $\sim$420 MeV. Moreover, extraction of GT strengths when the unit cross
section can not be calibrated using known strengths from $\beta$-decay, at present has poorly known
errors.

To provide the necessary systematics, accurate absolute differential cross sections
at forward scattering angles for transitions for which the Fermi or GT strengths are
known were obtained in two experiments performed at RCNP. In the first experiment,
the ($^{3}$He,$t$) reaction on $^{12}$C, $^{13}$C, $^{18}$O, $^{26}$Mg, $^{60}$Ni,
$^{90}$Zr, $^{120}$Sn and $^{208}$Pb was measured. The tritons were detected in the
focal plane of the Grand Raiden spectrometer \cite{FUJ99} up to laboratory
scattering angles of 2.5$^{\circ}$. The experiment was run in achromatic and
off-focus \cite{FUJ01} modes of operation and resolutions in excitation energy of
100 keV (FWHM) and scattering angle of 0.2$^{\circ}$ were achieved (for details, see
Ref. \cite{ZEG06} in which the $^{26}$Mg($^{3}$He,t) data are discussed in detail).
The systematic uncertainty in absolute cross sections was about 10\%, mainly due to
uncertainties in the beam-current integration using a Faraday cup.

\begin{table}
\caption{\label{tab:table1} Transitions to final states included in the analysis of
the GT and Fermi unit cross sections. The $B(GT)$ values were calculated from known
log$ft$ values \cite{ENSDF} following Ref. \cite{BRO93}.  For the Fermi strengths,
$B(F)=N-Z$ was used.}

\begin{ruledtabular}
\begin{tabular}{llll}
\multicolumn{2}{c}{GT} &\multicolumn{2}{c}{Fermi} \\
\hline ($J^{\pi}$, $E_{x}$ [MeV]) & B(GT) & ($J^{\pi}$, $E_{x}$ [MeV]) & B(F) \\
\hline
$^{12}$N(1$^{+}$, 0.) & 0.88 & $^{13}$N($\frac{1}{2}^{-}$, 0.) & 1 \\
$^{13}$N($\frac{3}{2}^{-}$, 15.1 ) & 0.23 & $^{26}$Al(0$^{+}$, 0.228) & 2\\
$^{18}$F(1$^{+}$, 0.) & 3.11 & $^{58}$Cu(0$^{+}$, 0.203) & 2 \\
$^{26}$Al(1$^{+}$, 1.06) & 1.10 & $^{60}$Cu(0$^{+}$, 2.54) & 4  \\
$^{58}$Cu(1$^{+}$, 0.) & 0.155 & $^{90}$Zr(0$^{+}$, 5.01)& 10 \\
$^{118}$Sb(1$^{+}$, 0.) & 0.344 & $^{118}$Sb(0$^{+}$, 9.3) & 18 \\
$^{120}$Sb(1$^{+}$, 0.) & 0.345 & $^{120}$Sb(0$^{+}$, 10.2) & 20 \\
                            &       & $^{208}$Bi(0$^{+}$, 15.1) & 44 \\
\end{tabular}
\end{ruledtabular}
\end{table}

\begin{figure}
\includegraphics[width=7.5cm]{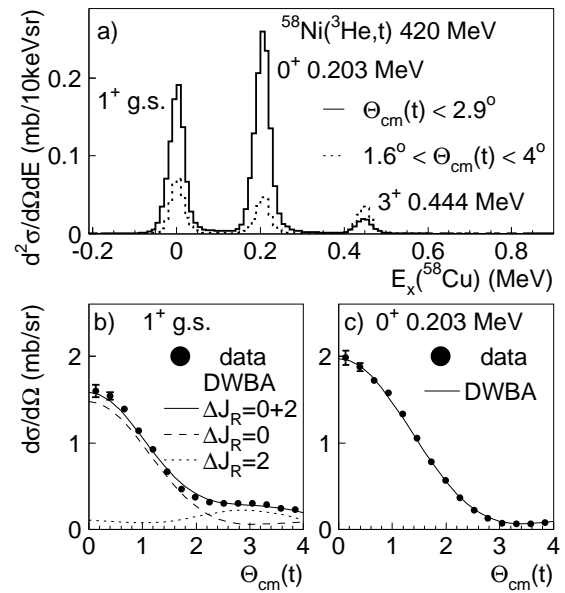}
\caption{\label{prl1}a) Low-energy part of the excitation-energy spectrum in
$^{58}$Cu measured via $^{58}$Ni($^{3}$He,$t$) at 420 MeV. The full line (dashed
line) corresponds to the spectrum taken with the spectrometer set at 0$^{\circ}$
(2.5$^{\circ}$). b) Measured differential cross section for the GT transition to the
$1^{+}$ ground state and fit with DWBA angular distribution. c) Idem for the Fermi
transition to the $0^{+}$ state at 0.203 MeV. Error bars are statistical only and
mostly smaller than the dots.}
\end{figure}

In the second experiment, cross sections on $^{26}$Mg, $^{58}$Ni and $^{118}$Sn were measured up to
4$^{\circ}$. The experiment was performed in dispersion-matched mode \cite{FUJ02}, resulting in
energy resolutions of $\sim$40 keV (FWHM). The $^{26}$Mg data were used to check for consistency
with the first experiment. Angular distributions measured in the two experiments were found to be
in good agreement, but due to inefficient current integration when running in dispersion-matched
mode, a 20\% correction had to be applied to all cross sections measured in the second experiment.

The differential cross sections were extracted for transitions to final states with
known GT and Fermi strengths listed in Table \ref{tab:table1}. For the case of
$^{13}$C, results from Ref. \cite{FUJ04} were also used in the analysis. The
extraction of the cross section for the excitation of the IAS in $^{13}$N required
special care, since the transition to the $\frac{1}{2}^{-}$ ground state contains
both GT and Fermi contributions. The GT contribution was removed using the GT unit
cross section extracted from the excitation of the $\frac{3}{2}^{-}$ state at 15.1
MeV, for which the B(GT) is determined from the $\beta^{-}$-decay of its isospin
multiplet partner $^{13}$B \cite{TAD87}.

As an example of the procedure for extracting unit cross sections, the results for the Fermi and GT
transitions from $^{58}$Ni to $^{58}$Cu are shown in Fig. \ref{prl1}. Figure \ref{prl1}a shows the
part of the excitation-energy spectrum that includes the GT transition of known strength to the
1$^{+}$ ground state and the Fermi transition to the 0$^{+}$ state at 0.203 MeV. Their differential
cross sections are plotted in Figs. \ref{prl1}b and \ref{prl1}c, respectively. To extract the cross
section at 0$^{\circ}$, the experimental differential cross sections are fitted to theoretical
ones, calculated in distorted-wave Born approximation (DWBA) using the code FOLD \cite{FOLD}.

For the optical potentials, parameters that extracted from $^{3}$He elastic scattering data
\cite{YAM95,KAM03} were used (the data from Ref. \cite{YAM95} were refitted). The effective
nucleon-nucleon interaction of Love and Franey \cite{LOV80} is double-folded over the transition
densities of the $^{3}$He-$^{3}$H and target-residual systems. For $^{3}$He and $^{3}$H, densities
were obtained from Variational Monte-Carlo results \cite{WIR05}. For the target-residual system,
one-body transition densities (OBTDs) were calculated with the shell-model code \textsc{OXBASH}
\cite{OXBA} and obtained from Ref. \cite{HON05}, using appropriate interactions for the nuclei in
Table \ref{tab:table1}. For Fermi transitions in nuclei heavier than $^{26}$Mg, OBTDs were
determined from a normal-modes procedure \cite{NOR}. Details on the DWBA calculations can be found
in Refs. \cite{ZEG06,COL06} and will be discussed in a forthcoming publication.

In the case of Fermi transitions (Fig. \ref{prl1}c) a single fit parameter was used to scale the
theory to the data. The extracted $0^{\circ}$ cross section was then extrapolated to $q=0$ using
the ratio of calculated cross sections in DWBA at $q=0$ and $0^{\circ}$. Of the transitions (Fermi
and GT) studied here, this ratio was maximally 1.25. For the GT transitions, a similar procedure
was used. However, the analysis is complicated by contributions from incoherent and coherent
$\Delta$L=2, $\Delta$S=1 contributions to the $\Delta$J=1 GT excitation. The incoherent
contribution, due to a transition in which the total angular momentum transfer ($\Delta$J$_{R}$) to
the relative motion between the target and projectile equals 2, was removed by fitting the angular
distribution to a linear combination of the theoretical $\Delta$J$_{R}$=0 and $\Delta$J$_{R}$=2
components calculated in DWBA, as shown in Fig. \ref{prl1}b. Only the $\Delta$J$_{R}$=0 component
is used to determine the $0^{\circ}$ GT cross section. The coherent contribution, largely due to
the non-central tensor-interaction and the largest source that breaks the proportionality of Eq.
\ref{eq:eik} \cite{ZEG06,COL06,FUJ07}, cannot easily be determined from the data since it does not
strongly affect the angular distribution at forward scattering angles. We estimated its effect on
the cross sections using the ratio of the DWBA calculations with and without the tensor force. Such
estimates have proven to provide reasonable predictions for the proportionality breaking in
$^{26}$Mg \cite{ZEG06} and $^{58}$Ni \cite{COL06}. For two of the cases studied here, effects
larger than statistical errors were predicted. For the excitation of the $^{58}$Cu($1^{+}$) state,
the cross section decreased by 20\% if the tensor force was excluded from the calculation. To
correct for this, the GT unit cross section extracted from the data should be reduced by 20\%
\cite{COL06}. For the GT component of the excitation of the $^{13}$N ground state (needed for
extraction of the Fermi strength in this transition, see above), the cross section decreased by
15\% if the tensor force was excluded. Its cross section estimate, based on the unit cross section
extracted from the 15.1 MeV state in $^{13}$N, must thus be increased by that percentage before
subtracting it from the total ground-state cross section for the purpose of deducing the cross
section of the Fermi component. Consequently, the Fermi unit cross section decreases. We initially
ignore these corrections and then show how they affect the relevant unit cross sections.

\begin{figure}
\includegraphics[width=7.5cm]{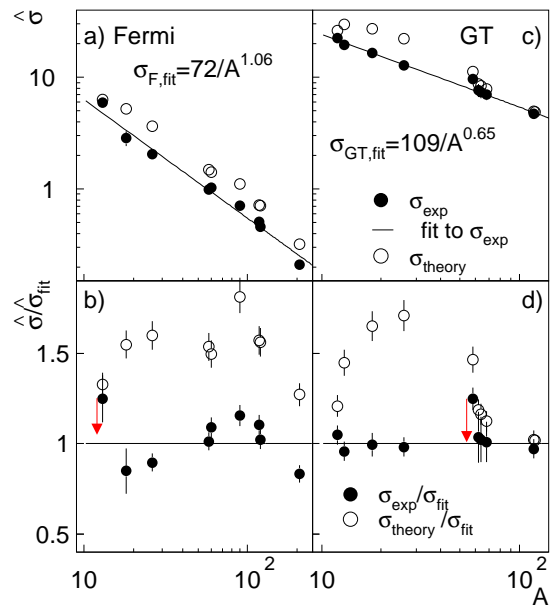}
\caption{\label{prl2}a) Mass dependence of the experimental and theoretical Fermi
unit cross sections. The line indicates a fit to the experimental values. b)
Relative deviation of the experimental and theoretical Fermi unit cross sections
from the fitted mass dependence in a). c) As in a), but for the GT unit cross
sections. d) As in b), but for the GT unit cross sections. The arrows in b)
(for$^{13}$C)  and d) (for $^{58}$Ni)  indicate to which value these unit cross
section change after theoretically estimated corrections are applied (see text). }
\end{figure}

The unit cross sections $\hat{\sigma}_{GT}$ and $\hat{\sigma}_{F}$ are calculated by
dividing the extracted cross sections at q=0 by the known $B(GT)$ and $B(F)$ values,
respectively. In Fig. \ref{prl2}a the empirical Fermi unit cross sections are
plotted against mass number. A very smooth decreasing trend is seen which is well
fitted with the function $\hat{\sigma}_{F,fit}=72/A^{1.06}$. By dividing the
measured Fermi unit cross sections by $\hat{\sigma}_{F,fit}$ (Fig. \ref{prl2}b) it
is seen that the experimental unit cross sections deviate by no more than 15\% from
the fit, except for the case of the IAS in $^{13}$N. After the above-mentioned
correction to GT component of this transition, the Fermi unit cross section for this
nucleus decreases by an amount indicated by the arrow in Fig. \ref{prl2}b and its
corrected value is close to the trend line.

In both Figs. \ref{prl2}a and b, the theoretical unit cross section calculated by
dividing the DWBA cross sections at $q=0$ for each transition by $B(F)=N-Z$ are also
shown. The trend follows the data, but the values are 30\%-80\% too high. Further
development of the reaction codes is needed to understand this discrepancy, but
likely sources are the approximate treatment of exchange \cite{UDA87}, density
dependence of the nucleon-nucleon interaction and uncertainties in the optical
potentials.

In Figs. \ref{prl2}c and d, similar plots are shown for the GT unit cross sections.
To complement the results, GT unit cross sections for A=62, 64 and 68 were
calculated by multiplying measured $\frac{\hat{\sigma}_{GT}}{\hat{\sigma}_{F}}$
ratios \cite{ADA06} by the $\hat{\sigma}_{F,fit}$ fit function described above. The
mass number dependence of the GT unit cross section is well fitted with the function
$\hat{\sigma}_{GT,fit}=109/A^{0.65}$ (Fig. \ref{prl2}c) and the measured GT unit
cross sections deviate by no more than 5\% from this function, except for $^{58}$Cu.
After the above-mentioned correction, the unit cross section for this case reduces
(see arrow in Fig. \ref{prl2}d) to a value close to the trend line. The theoretical
GT unit cross sections, calculated by dividing the $\Delta$L=0 DWBA cross sections
at $q=0$ for each transition by its corresponding theoretical $B(GT)$ overestimate
the data by up to 70\%, with the largest deviations seen for medium-mass nuclei.

It is important to note that the $\tau$ component of the nucleon-nucleon interaction is short-range
in nature, whereas the $\sigma\tau$ component is dominated by long-range terms \cite{LOV80}. The
difference in range between the two is responsible for the different dependences of GT and Fermi
unit cross sections (and thus their ratio \cite{ADA07}) on mass number and this is seen in both
theory and data. The long-range nature of the $\sigma\tau$ interaction partially counters the
strong absorption of the probe on the surface of the nucleus and this effect becomes stronger for
heavier nuclei. The ($p,n$) reaction is different in this regard since it probes the interior of
the nucleus. Consequently, the difference in range between $\tau$ and $\sigma\tau$ components
matters less and the empirical target mass dependences of Fermi and GT unit cross sections are very
similar \cite{TAD87}.

In summary, we have analyzed absolute differential cross sections over a wide mass range and found
that the empirical mass dependences of Fermi and GT unit cross sections for the ($^{3}$He,$t$)
reaction at 420 MeV are well described by simple relationships. This puts the use of this probe to
extract weak transition strengths on a solid phenomenological footing and makes it possible to
extract with confidence GT strengths in nuclei for which the unit cross section cannot be
calibrated using transitions with known B(GT) from $\beta$-decay. The theoretical unit cross
sections calculated in distorted-wave Born approximation overestimate the data but are adequate to
estimate corrections to GT cross sections due to the non-central tensor component of the
interaction and qualitatively explain the difference in mass dependence between Fermi and GT unit
cross sections.

\begin{acknowledgments}
We wish to express our gratitude to the cyclotron staff at RCNP. This work was supported by the US
NSF (PHY-0216783 (JINA), PHY-0606007 and PHY-0555366), the Ministry of Education, Science, Sports
and Culture of Japan, the Stichting voor Fundamenteel Onderzoek der Materie (FOM), the Netherlands
and the DFG, under contract Br 799/12-1.
\end{acknowledgments}

\bibliography{prl}

\end{document}